# Towards Privacy Engineering for Real-Time Analytics in the Human-Centered Internet of Things

Thomas Plagemann, Vera Goebel
University of Oslo
Norway

Matthias Hollick
Technical University of Darmstadt
Germany

Boris Koldehofe
University of Groningen
The Netherlands

**Foreword**

It is the aim of this white paper to give the interested reader an insight into the motivation, background, research challenges, and approaches of the Parrot research project (Privacy Engineering for Real-Time Analytics in Human-Centered Internet of Things), funded by the Research Council of Norway (2021 – 2025, Project number 311197).

For recent information about the project, please refer to its home page: https://www.mn.uio.no/ifi/english/research/projects/parrot/

## Abstract

Big data applications offer smart solutions to many urgent societal challenges, such as health care, traffic coordination, energy management, etc. The basic premise for these applications is "the more data the better". The focus often lies on sensing infrastructures in the public realm that produce an ever-increasing amount of data. Yet, any smartphone and smartwatch owner could be a continuous source of valuable data and contribute to many useful big data applications. However, such data can reveal a lot of sensitive information, like the current location or the heart rate of the owner of such devices. Protection of personal data is important in our society and for example manifested in the EU General Data Protection Regulation (GDPR). However, privacy protection and useful big data applications are hard to bring together, particularly in the human-centered IoT. Implementing proper privacy protection requires skills that are typically not in the focus of data analysts and big data developers. Thus, many individuals tend to share none of their data if in doubt whether it will be properly protected. There exist excellent privacy solutions between the "all or nothing" approach. For example, instead of continuously publishing the current location of individuals one might aggregate this data and only publish information of how many individuals are in a certain area of the city. Thus, personal data is not revealed, while useful information for certain applications like traffic coordination is retained.

The goal of the Parrot project is to provide tools for real-time data analysis applications that leverage this "middle ground". Data analysts should only be required to specify their data needs, and end-users can select the privacy requirements for their data as well as the applications and end-users they want to share their data with. The project results are expected to enable the (semi-)automatic integration of appropriate privacy protection into real-time data stream applications. Thus, individuals can safely provide data which in turn improves the results of big data applications.

## 1. Introduction

Sensor technology and data analytics in the Internet of Things (IoT) change the way we interact with the physical world and enable fast data-driven decision-making. This fundamentally transforms processes and systems in all industry sectors, health, traffic, city management, etc., and promises increased efficiency, safety, and better health. McKinsey analyzed the potential economic impact of IoT applications and estimates it of as much as $11.1 trillion per year in 2025 for nine selected domains [1]. The value of IoT applications comes from analyzing data from many sensors for software-based decision-making. As such, sensor data and software to analyze the data play a key role as IoT enablers, but also as a potential barrier. One important reason for the barrier is the fact that implementation of proper privacy protection in IoT applications is a very challenging engineering task and requires expertise in security, privacy, and data analytics; and privacy protection is often

not properly implemented. However, the oxymoron of privacy protection in big data is caused by the fact that missing privacy protection finally leads data subjects to provide no data or data of very low quality [2].

Parrot aims to diminish this barrier with a privacy engineering solution with efficient system support for privacy protection. We exploit as a use-case the domain of human-centered IoT in which individuals (i.e., data subjects) provide sensor data from their smartphones and other devices; and we focus on real-time analytics to enable the fastest possible decision-making. The conflict between data subjects and data analytics is omnipresent in this domain [3]. Informed data subjects want to protect as much of their privacy as possible, but increasing the quantity and quality of data enables better results for decision-making. One core insight for this project is that privacy is not restricted to the *all-or-nothing* policy, i.e., either there is access to all sensor data or no data at all. Instead, there are several *levels* of privacy protection in between. These levels might be acceptable for data subjects and provide data (with reduced quality) that is still useful for data analytics. Obfuscation [4] and k-anonymity [5] are two good examples of implementing such levels. The resulting research challenge is how to match the privacy requirements of data subjects with the requirements of data analytics.

The second core insight is that privacy requirements are not static and depend on the use of the data in terms of applications, end-users, threat models, and the context of the data subject [6], [7]. It should be noted that one sensor data stream can be used at the same time for different applications, which might require different levels of privacy protection. Consider the simple example of location data: a data subject might be interested in always providing the exact GPS coordinates to near family members, but to healthcare personnel only if there is a certain probability that a critical health condition might happen. Therefore, data subjects need to be enabled to express privacy requirements for their data and the particular use of it (i.e., who uses it for what). The resulting research challenge is to address the lack of a unified approach to *quantify* the level of privacy protection that can be achieved with different privacy-protecting mechanisms (PPM) [9] to enforce the privacy requirements, and real-time adaptation of privacy protection to the context of individual data subjects.

The third core insight is that *privacy by design* requires considering privacy during the entire engineering process, but it does not prohibit the separation of concerns (e.g., privacy, efficiency, and data analytics) as an engineering principle to allow application developers to focus on the application logic. However, there is a strong dependency between data analytics and privacy since PPM often degrade the utility of data for the analytics. Therefore, we propose to address this research challenge by a declarative approach (i.e., data subjects and application developers declare their privacy and data quality requirements) and "*automated enforcement of privacy requirements and preferences*"[1] as recently recommended by the European Union Agency for Network and Information Security (`enisa`).

The data produced by IoT devices is almost by definition streaming data, i.e., data samples are continuously generated, typically with high velocity and at high volumes. Furthermore, for many applications, there is the demand for analyzing the data as soon as possible after its production, i.e., to perform stream analytics. Classical data management platforms for data analytics, like data warehouses, are not well-suited for this demand. Therefore, data streaming processing solutions like Complex Event Processing (CEP) are a central part of most recent data analytics platforms. The fact that many of the global IT players provide corresponding products, like IBM, Microsoft, SAP, AMAZON, Google, etc., indicates how well data stream processing is adopted in industry and academia. CEP can stepwise transform basic events from multiple sources into complex events by detecting patterns such as causal and temporal relations over the event streams. Machine Learning (ML) could be used together with CEP, but it cannot substitute CEP. As a statistical approach ML cannot guarantee correctness, is not yet suited to detect complex patterns over multiple data streams, and for many application domains sufficient training data is not available. However, privacy protection in data stream processing faces many open challenges [10], and privacy in CEP is in its infancy as a research area [11], [12]. The classical processing model in data analytics is to transport all data from the sources to a data center. However, the closer data aggregation and privacy protection is performed to the sources, the harder it is for attackers to breach security and privacy protection. As such, privacy protection can benefit from moving from the cloud-based approach to fog computing, or even to fully distributed in-network processing. Additionally, processing close to the data sources can substantially reduce resource consumption in terms of energy, bandwidth, etc.; and allow for sophisticated load balancing solutions. However, the support of different distribution levels over different networks, like 5G and device-to-device communication, is still a large challenge. It is necessary to develop new solutions for automatic distribution, and to analyze their advantages and disadvantages in terms of costs, efficiency, and vulnerability. This insight is in line with the `enisa` recommendation "*the research community and the big data analytics industry need to continue and combine* *their efforts towards decentralized privacy preserving analytics models*" [2]. Therefore, an adaptive Distributed CEP (DCEP) solution for real-time analytics and privacy protection is subject of our research.

## 2. Use cases

We will introduce the concept of Smart Communities to support tailored privacy protection for a group of data subjects that serve one instance of an IoT application and have the same privacy requirements. To demonstrate the need for tailored privacy protection in Smart Communities we use three different IoT applications that all use heart rate, GPS, and accelerometer sensor data. A mHealth application analyzes heart rate variability and activity to estimate the risk for cardiac morbidity and mortality; and track health changes due to medication or life-style changes [8]. In a fitness application, the data is used to analyze and share the activity and fitness level with friends; and in a city stress map application, density and movement of individuals shall be visualized together with stress indicators, e.g., to avoid respectively manage stressed crowds in the city. Identity of data subjects, location, and physiological data require different privacy protection levels in these applications. Location privacy is by default supported in the mHealth application, but if a critical health situation is detected location information is revealed to enable quick intervention. Obfuscation and data aggregation could be applied to provide some privacy level when sharing fitness data with friends, and k-anonymity could protect the privacy of individuals and still deliver useful information for the city stress map.

Figure 1 illustrates a Smart Community for the fitness application in which the real-time analysis of the sensor data from four data subjects is done in the DCEP instance $DCEP_F$. Furthermore, it shows that three Smart Communities are supported on one device that are separated through a platform called DevCom [27]. DevCom provides a decentralized VPN-like[1] protection and nodes can be in multiple "VPNs" (aka Smart Communities) at the same time. Each Smart Community maintains a particular level of privacy protection determined by the privacy requirements of data subjects and data quality requirements of the application.

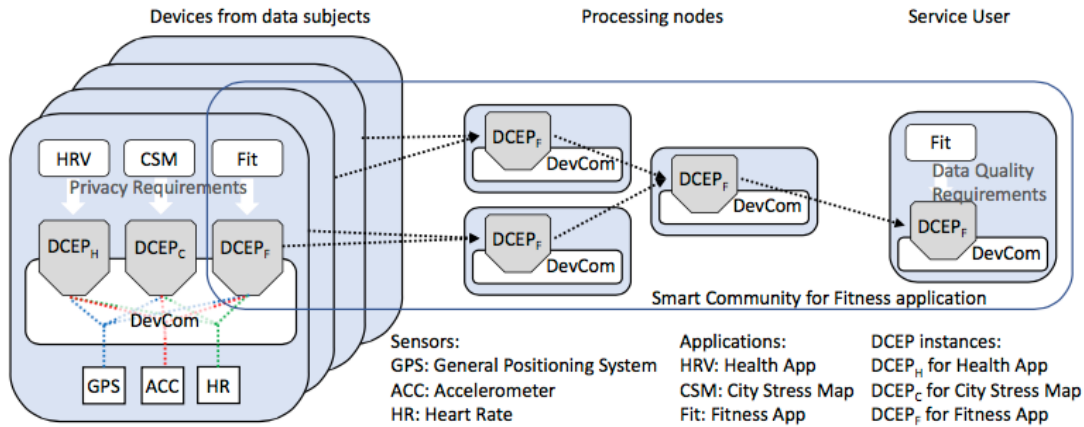

Figure 1: Structural view of a Smart Community

## 3. State-of-the-Art

In this section, we briefly discuss the state-of-the-art related to our main contributions: (1) privacy engineering tools to automatically enforce privacy in applications, (2) privacy support in adaptable CEP and IoT middleware, (3) specification of privacy requirements and preferences, quantification of privacy protection, and data quality needs, and (4) adaptive privacy protection. Furthermore, we summarize the state of privacy in participatory sensing, because it is probably the most intensively investigated human-centered IoT area.

(1) Privacy engineering tools to automate tasks like privacy impact assessment [13], data mapping to maintain records of data processing activities [14], and discovery and mapping of personal information [15] exists, but enhancing IoT applications automatically with privacy enforcement is an open challenge.

---
[1] VPN: Virtual Private Network

(2) Privacy support in CEP and IoT middleware: There exists very little work on privacy protection in CEP. Probably the first publication on this topic is a theoretical analysis of the trade-off between quality and privacy in CEP [11]. Wang et al. [12] present a practical solution for privacy-preserving CEP through event suppression. PeGaSus [16] is an algorithm to answer continuous queries over data streams under differential privacy. Schilling et al. [17] combine fine-grained access control with obfuscation to increase the utility of the privacy-preserving CEP system. There exist more work on privacy protection in the general area of data stream processing and publish-subscribe systems. Many of them use access control, which basically implements an *all-or-nothing* policy. The IoT middleware developed in the RERUM project [21] shares probably the most commonalities with this proposal since it targets security and privacy by design through a middleware that uses CEP to process sensor data. However, it is a static approach that provides anonymization through the removal of IDs and access control. Privacy quantification and dynamic configuration of privacy support is not considered. A recent survey on IoT middleware concludes that of the investigated solutions "*none support privacy-by-design and trust management*" [22]. This is a rather strong conclusion but can be seen as an indication that research in this area is still at an early stage.

(3) Privacy specification, quantification, and data utility: In [18], a middleware solution is presented that maps services to personal information types with the help of ontologies to control the access to personal information. A markup language is introduced in [19] to represent user preferences and privacy policies and is used in a middleware solution to support privacy in context-aware applications. The principles of Service Oriented Architectures are applied to design a service-oriented middleware for privacy protection in pervasive computing in [20]. Privacy protection is achieved through a particular service access protocol. The survey of quantification of privacy preserving data mining algorithms demonstrates well that a unified solution for privacy quantification is missing [23]. Dasgupta et al. [24] use a model and taxonomy of visual uncertainty to define metrics for privacy and utility in privacy preserving visualization. The major conclusion in [25] is that comparing the effect of anonymization-based privacy protection on information quality for data mining is an unsolved issue. Very recently, Wagner et al. [25] published a systematic survey of technical privacy metrics describing 80 different metrics. This documents very well that there is no solution towards a unified privacy quantification.

(4) Adaptive privacy protection: Most context-aware privacy protection solutions use access control to prohibit access to sensor data in certain contexts. The Thesis presented in [37] gives a good overview over these approaches. Olejniki et al. [38] go a step further and provide the possibility to obfuscate the data as alternative to access allowed or access denied. Garcia-Morchon et al. [39] extent role-based access control systems with access control decisions that are based on the health acuteness of a user. Situation-aware access control is combined in [40] with purpose-oriented situation models.

Privacy in participatory sensing is still a big and largely unsolved challenge. A good and still valid overview on these challenges is given by Shilton and Estrin in [30]. Privacy-preserving data aggregation is addressed in [33] and [34]. PEPSI [35] is a recent solution that is based on Identity-Based Encryption to achieve a privacy enabled participatory sensing infrastructure. Important works for our goals related to privacy include the frameworks for personalized privacy [36] and formal semantics of privacy protection in [41]. Surveys on privacy in mobile participatory sensing applications are given in [31] and [32]. The study in [9] is a recent update of [32] and concludes with the identification of several open challenges that are actually addressed in this proposal: "*providing composable privacy solutions*", "*trade-offs between privacy, performance, and data fidelity*", "*making privacy measurable*" and "*holistic architecture blueprints*".

## 4. Research questions and approach

The research in the Parrot project is driven by the following three research questions (RQ):

RQ1: Which formalisms and knowledge representations are suitable to describe the different aspects of "proper privacy protection", i.e., to (1) specify privacy requirements of data subjects, privacy policies of smart communities, and data quality requirements of data analytics, (2) model how PPM impact data quality, and (3) quantify the level of privacy protection that can be achieved by different PPM and a given threat model?

RQ2: Can such knowledge representations can be used (and how) to rewrite CEP queries automatically and configure DCEP in a Smart Community such that (1) privacy requirements of data subjects and privacy policies of the Smart Community are enforced, (2) data quality requirements of data analytics are met,

(3) costs of data acquisition and processing are minimized, (4) costs of potential privacy attacks are maximized, and (5) the provided service enables better real-time decision making for service users. In other words, an application developer focuses on the application logic and proper privacy protection is automatically added?

RQ3: Can Event Proximity[2] [43] be used to achieve dynamic adaptation of the privacy protection level to the context of individual data subjects, e.g., their health situation?

The methods used to investigate these RQs comprise the definition of proper means for knowledge representation, the use of analytic and semantic models to solve two optimization problems, the design and implementation of adaptive DCEP for Smart Communities and IoT applications, and quantitative and qualitative evaluation through simulation and deployment and field tests in an international testbed. In the following paragraphs, we briefly outline how these research methods (RM) will be used.

**Knowledge representation**: Automatically selecting the most suitable PPM for a given CEP query requires gathering and representing knowledge depending on (1) which level of privacy protection existing PPM can achieve given a particular threat model, and (2) the impact of PPM on data quality. There are two basic approaches for knowledge representation: symbolic (e.g., expert systems and ontologies) and non-symbolic (e.g., neural networks in ML). The latter has the disadvantage that humans are not able to interpret this knowledge. Still, we aim to make the gained knowledge available for developers that do not use our software or even CEP. Given the maturity and tools available, we select to use an ontology. Furthermore, specification languages and respective frameworks are needed for dealing with privacy policies and data quality requirements.

**Mathematical optimization**: The two tasks that require solutions for multi-dimensional optimization problems are (1) the selection of privacy protection mechanisms for CEP query rewriting (Figure 2a) and (2) the generation and placement of processing graphs, also called operator placement (Figure 2b). We envisage following a utility-based approach to these problems. The utility of a system configuration is usually approximated by a convex utility function over the properties of the system and the system context. For the selection of privacy protection mechanisms, the utility of the selection is determined by the usefulness of the service. The usefulness is expressed in terms of achievable Quality-of-Information and Quality-of-Experience for the end-user under the constraints that the privacy requirements of the data subject and privacy policies of smart communities are fulfilled and the data quality requirements of the data analytics are met. The utility of placed processing graphs is defined by the potential costs incurred through communication and processing (measured in terms of resource consumption and monetary costs), and how vulnerable the configuration is to privacy attacks. PPM can introduce constraints for the placement, e.g., k-anonymity requires to process data streams from at least k data subjects. Input to solve the optimization problems comprises the definition of the privacy requirements, privacy policies, privacy quantification, characteristics of available PPM, the quality requirements of the data analytics as well as information about the available infrastructure and costs. Since the combinatorial complexity might be high, it is reasonable to assume that heuristics are needed to achieve the necessary performance. Another challenge is to rewrite CEP queries automatically to integrate the selected PPM. One important input to this rewriting task is an ontology describing PPM.

---

[2] In CEP systems, an event of interest is either detected or not. Event proximity complements the classical discrete set of possible results with a a continuous result domain and quantifies the closeness of the current situation to an event of interest. Thus, it enables the system and its users to react to anomalies and hazards before they actually happen.

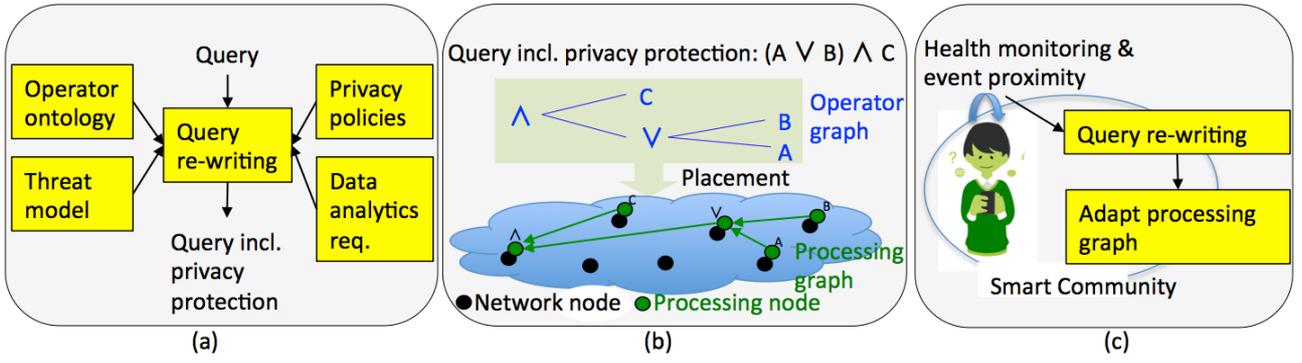

Figure 2: (a) Query rewriting, (b) Middleware adaptation, (c) Health aware privacy protection

**Design and implementation** of adaptive DCEP for Smart Communities and IoT applications is an important method to address systems research challenges and to transfer the conceptual and analytical research results to solutions that can be deployed and demonstrated in real systems. VPNs are the state-of-the-art to provide a trustworthy overlay for a user. However, a device can at any time be only in one VPN. To enable a device to be a member of different trustworthy Smart Communities at the same time, we need to apply concepts from DevCom [27] and add sensor support and hooks to integrate the adaptive DCEP. The DCEP will be responsible for data acquisition and real-time data processing. For run-time adaptability, we use CEP to detect events that might trigger adaption. Query rewriting and operator placement will be used to determine whether privacy mechanisms and the processing graph should be changed. We adopt models, mechanisms, and engineering approaches of dynamic software product lines [44] to perform architectural adaptation in form of operator selection and placement. The result of the analysis and planning phase is a processing graph (see Figure 2b), which is instantiated and executed by the DCEP. Furthermore, DCEP needs to perform monitoring to provide data about the available infrastructure and resources for operator placement. For field tests and deployment, a component-based DCEP design will be used to develop an adaptive DCEP for Linux and Android platforms.

**Quantitative and qualitative evaluation**: The query rewriting and the adaptive DCEP need to be analyzed to understand their performance, costs, and overheads for deployment strategies, like centralized solutions, fog processing, and fully distributed in-network processing, in different settings in terms of the number of sensor streams and data subjects, mobility and density of nodes, etc. Such evaluations will be performed as early as the first functional parts are implemented and will be studied first in a sandbox environment. Later on, network simulations, laboratory experiments with the real implementation, and deployment on a RaspberryPi and smartphone/Android infrastructure. The deployment will be used for measurements during the field test. Design and implementation of the IoT applications for the field test will be used to gain quantitative evaluation results related to the development process.

## 5. Impact

To describe the benefits for the society, we first identify the direct impact of the Parrot project by describing how all stakeholders in the value chain of IoT systems and applications (Figure 3) will benefit from the project results.

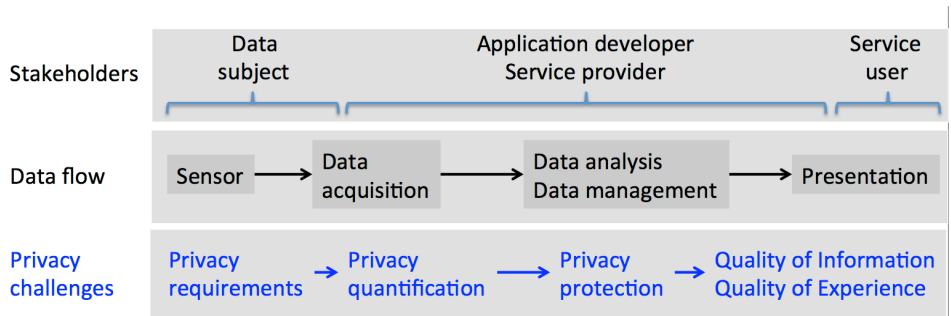

Figure 3: Stakeholders, data flow, and privacy challenges

**Application developers**: Automatic query rewriting relieves application developers from implementing privacy protection and the adaptive DCEP relieves them from considering deployment issues, like placement of data processing. Thus, they can focus all their creative energy on the development of new and better applications, which in turn can be reflected in increasing financial revenue. Knowledge gained on privacy quantification, relationship between privacy requirements and PPM, and the impact of PPM will also simplify the task of developers that do not rely on CEP.

**Service providers**: Adaptive DCEP reduces the effort of setting up and maintaining the infrastructure to run the application and provide the service. The evaluation of different deployment strategies in different settings will provide important knowledge to select the best deployment strategy. Consequently, the service can be provided with a protection level at lower cost, being more competitive, and having higher economic incentives.

**Data subjects**: Data subjects will not experience the dilemma to decide whether to provide all or no data, instead they can rely on fine-grained levels of privacy protection. Furthermore, the privacy level can be adapted in real-time to the context (e.g., health status) of data subjects such they can receive immediate help if needed.

**Service users**: We expect more data subjects to provide more data with higher quality because of the fine-gained levels of privacy enforcements. This increased quantity and quality of data means better services and a better foundation for decision making which shall ultimately lead to better decisions.

**Researchers**: The metric for privacy quantification will help researchers to compare PPM and thus increase the credibility of new research results. The open source adaptive DCEP allows to easily integrate arbitrary privacy protection mechanisms as new operators; and to test, evaluate and demonstrate them in simulation and real-world prototypes. This substantially increases the efficiency of development, testing, evaluation, and demonstration of new PPM. The measurement and tracing data collected during deployment and field tests will be provided as open datasets to the research community, which in turn will be excellent input for future research activities.

The Parrot project results shall make it easy to avoid the oxymoron of privacy protection and big data, and lead to increasing quantity and quality of data. A larger quantity and higher quality of data will in turn lead to better data analysis and real-time decision-making. Systems for real-time decision support will play an increasing role in many areas, like the economy, health care, smart cities, industrial production, environment management, etc. Obviously, better decision-making will lead to stronger improvements in all these areas.